\documentclass[pra, showkeys, reprint, nofootinbib]{revtex4-1}
\usepackage[english]{babel}
\usepackage[utf8]{inputenc}
\usepackage[colorinlistoftodos, color=green!40, prependcaption]{todonotes}
\usepackage{amsthm}
\usepackage{mathtools}
\usepackage{amsmath}
\usepackage{physics} 
\usepackage{graphicx}
\usepackage{soul}
\usepackage{bm}
\usepackage{mathbbol}

\newtheorem{definition}{Definition}

\newtheorem{prop}{Proposition}

\begin{document}
\title{Training Saturation in Layerwise Quantum Approximate Optimisation}

\author{E.~Campos} \email[e-mail:~]{ernesto.campos@skoltech.ru} \homepage{http://quantum.skoltech.ru}
    \affiliation{Skolkovo Institute of Science and Technology, 3 Nobel Street, Moscow, Russia 121205}
\author{D.~Rabinovich}    \email[e-mail:~]{daniil.rabinovich@skoltech.ru}
    \affiliation{Skolkovo Institute of Science and Technology, 3 Nobel Street, Moscow, Russia 121205}
\author{V.~Akshay} \email[e-mail:~]{akshay.vishwanathan@skoltech.ru}
    \affiliation{Skolkovo Institute of Science and Technology, 3 Nobel Street, Moscow, Russia 121205}
\author{J.~Biamonte}  \email[e-mail:~]{j.biamonte@skoltech.ru}
    \affiliation{Skolkovo Institute of Science and Technology, 3 Nobel Street, Moscow, Russia 121205}
   
\date{June 2021}

\begin{abstract}
 
Quantum Approximate Optimisation (QAOA) is the most studied gate based variational quantum algorithm today.  We train QAOA one layer at a time to maximize overlap with an $n$ qubit target state. Doing so we discovered that such training always saturates---called \textit{training saturation}---at some depth $p^*$, meaning that past a certain depth, overlap can not be improved by adding subsequent layers.  We formulate necessary conditions for saturation. Numerically, we find layerwise QAOA reaches its maximum overlap at depth $p^*=n$. The addition of coherent dephasing errors to training removes saturation, recovering robustness to layerwise training. This study sheds new light on the performance limitations and prospects of QAOA.

\end{abstract}

\maketitle

\section{Introduction}

Variational quantum algorithms are the centerpiece of study in modern quantum computing. Such algorithms are designed to alleviate certain systematic limitations of near term devices, including variability in pulse timing and coherence limitations \cite{harrigan2021quantum,pagano2019quantum, guerreschi2019qaoa,butko2020understanding}, at the cost of classical outer loop optimization. In particular, the Quantum Approximate Optimization Algorithm (QAOA) \cite{Farhi2014} was developed to approximate solutions to combinatiorial optimization problems \cite{niu2019optimizing,Farhi2014,lloyd2018quantum,morales2020universality,Zhou2020,wang2020x,Brady2021,Farhi2016,Akshay2020,Farhi2019a,Wauters2020,Claes2021,Zhou}.
Recent milestones include experimental demonstration of $3p$-QAOA (depth-three, corresponding to six tunable parameters) using 23 qubits \cite{harrigan2021quantum}, universality results \cite{lloyd2018quantum,morales2020universality}, as well as several results that aid and improve on the original implementation of the algorithm \cite{Zhou2020,wang2020x,Brady2021,akshay2021parameter}. Although QAOA recovers optimal query complexity in Grover's search \cite{Jiang2017a} and offers a pathway towards quantum advantage \cite{Farhi2016}, limitations are known for low depth QAOA \cite{Akshay2020,hastings2019classical,Bravyi2019}. Exact analysis is scarce and only describes QAOA on specific instances including e.g.~fully connected graphs and projectors \cite{Farhi2019a, Wauters2020,Claes2021, akshay2021parameter}.

As to most variational algorithms, QAOA consists of an outer loop classical optimization which tunes parameters of a quantum circuit to minimize (or maximize) an objective function. Optimization becomes increasingly challenging by considering more parameters, which increase linearly with depth. Various techniques have been developed to aid in optimization---leveraging problem symmetries \cite{shaydulin2021exploiting}, and parameter concentrations \cite{streif2019comparison, akshay2021parameter}, which these same authors previously studied. Several heuristic strategies have been explored to speed-up this challenging outer loop optimization step.

Layerwise training, a greedy learning strategy employed to reduce optimization time, has been shown to function via a reparameterization of search parameters \cite{Zhou}, and to be helpful at avoiding barren plateaus \cite{skolik2021layerwise}. Although promising, such strategies can become sub-optimal in certain scenarios i.e.~when stacking single layers is not expressive enough and adding more layers per stack is required to minimize a cost function, see abrupt training transitions \cite{campos2020abrupt}.

We studied this layerwise training strategy, where the objective function is the overlap, maximized by the target state. We discovered the onset of saturation at depth $p = n$  with maximal optimization returning trivial optimal angles for subsequent layers. Hence we discovered the maximal point for layerwise training. We found that layerwise QAOA always saturates and we derive necessary conditions for the circuit output (Proposition \ref{prop_necesary_conditions}).

Our results demonstrate that local coherent phase errors sampled from a Gaussian distribution remove the effect of saturation, suggesting a means to avoid non-trainable states that would otherwise satisfy the necessary saturation conditions. This systematic error can be considered as a local coherent noise model \cite{bravyi2018correcting}. Several works have demonstrated noise resilience of variational algorithms \cite{sharma2020noise, gentini2020noise, cincio2021machine, mcclean2016theory, mcclean2017hybrid}, while \cite{cao2021noise} demonstrated performance benefits that noise can induce for training a quantum autoencoder.

The present paper defines layerwise QAOA in Sec. \ref{sec_state_preparation}. We introduce the definition of training saturation and numerically demonstrate  that layerwise training saturates in Sec. \ref{sec_saturation}. 
In Sec. \ref{nec_cond} we derive the necessary conditions satisfied by families of saturated states. Sec. \ref{sec_noise} presents strategies to mitigate the effects of saturation focusing on the use of coherent phase noise. We reflect on the obtained results in the Discussion section. 

\section{Layerwise QAOA}
\label{sec_state_preparation}
QAOA can be viewed as variational state preparation: let $\ket{\bm{t}}$ be the target state in the computational basis. The task is to variationaly prepare a candidate state with high overlap with  $\ket{\bm{t}}$. In QAOA, the candidate state $\ket{\psi_p (\bm\gamma, \bm\beta)}$---prepared by a $p$ depth circuit---is parametrized as:
\begin{equation}
    \ket{\psi_p(\bm\gamma,\bm\beta)} =  \prod\limits_{k=1}^p e^{-i \beta_k \mathcal{H}_{x}} e^{-i \gamma_k \ketbra{\bm{t}}{\bm{t}}}\ket{+}^{\otimes{n}},
    \label{ansatz}
\end{equation} 
with real parameters $\gamma_k\in[0,2\pi)$, $\beta_k\in[0,\pi)$. Here $\mathcal{H}_x = \sum_{i=1}^{n} X_{i}$ is the standard one-body mixer Hamiltonian with Pauli  matrix $X_i$ applied to the $i$-th qubit.

The optimization task maximizes the overlap between the candidate state $\ket{\psi_p(\bm\gamma,\bm\beta)}$ and the target state $\ket{\bm t}$ given by

\begin{equation}
0\le \max_{\bm \gamma, \bm \beta} \abs{\braket{\bm t}{\psi_p(\bm \gamma, \bm \beta)}}^2\le 1
\label{overlap}
\end{equation}

The global optimization strategy for fixed circuit depth $p$ is to maximize the overlap function \eqref{overlap} over all $2p$ variational parameters . The optimization task is evidently challenging for deep circuits. Several approaches have been suggested to simplify optimization including leveraging parameter concentrations \cite{akshay2021parameter,streif2019comparison} and exploiting problem symmetries \cite{shaydulin2021exploiting}. 

An alternative strategy suggests training circuits {\it layerwise} which reduces time, and to avoids barren plateaus \cite{skolik2021layerwise}. In this greedy approach, parameters of each layer are optimized one layer at a time: in each iteration a new layer is added, and only the new layer parameters are exclusively optimized (with all antecedent layers fixed as per previous iterations). The process is iterated subject to certain termination criteria (reaching a threshold overlap).

Whereas layerwise training significantly simplifies outer loop optimization, it necessitates deeper circuits compared to global optimization.

\section{Saturation}
\label{sec_saturation}
Here we consider numerical results of training layerwise QAOA. It was shown previously in \cite{akshay2021parameter} that the overlap is insensitive to the target state $\ket{\bm t}$, therefore without loss of generality we consider the case $\ket{\bm t}=\ket{0}^{\otimes n}$. 
\begin{definition} [Saturation]
Training saturates for depth $p^*$, the smallest depth for which 
\begin{equation}
   |\bra{t}\ket{\psi_{p^*+1}}|^2\leq|\bra{t}\ket{\psi_{p^*}}|^2\neq 1, 
\end{equation}
where the overlaps are maximized layerwise. 
\end{definition}

In other words, training saturates when an extra layer (and thus all subsequent layers) does not allow for overlap improvement. 

\begin{figure}[ht!]
    \centering
    \includegraphics[width=0.45\textwidth]{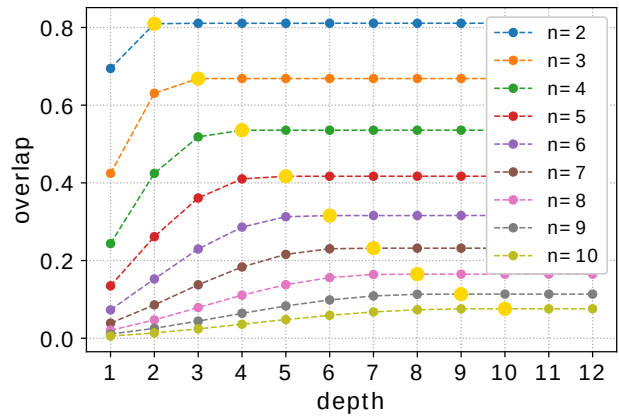}
    \caption{Saturation in layerwise training. Saturation occurs at $p^*=n$ as indicated by the highlighted points. }
    \label{saturability}
\end{figure}

Our results demonstrate layerwise training always saturates, which limits this method's performance compared to globally optimized QAOA. 
Specifically we observed an $n$-dependent threshold depth $p^*$ beyond which overlap cannot be improved further by optimizing additional layers. The induced saturation is also evident as the overlap for any new layer achieves optimality at trivial values of variational parameter $\beta_{p}$, leaving the circuit output state invariant when optimizing subsequent layers.
We find that saturation always occurs from the threshold depth $p^* = n$ (see Fig. \ref{saturability}).

Given the onset, this result appears to severely limit layerwise training as an approach towards alleviating the tedious computational cost suffered in global optimization.

Being a greedy algorithm, layerwise training, in contrast to global training, maximizes overlap increase from layer to layer. Therefore, we observe that for a set of globally optimized parameters $\{\bm{\gamma}^*, \bm{\beta}^*\}$ and a set of layerwise optimized parameters $\{\bm{\gamma}^\#, \bm{\beta}^\#\}$, there exists a certain depth $c$:
\begin {equation}
|\bra{0}\ket{\psi_c(\bm{\gamma}^*, \bm{\beta}^*)}|^2<|\bra{0}\ket{\psi_c(\bm{\gamma}^\#, \bm{\beta}^\#)}|^2,
\end{equation}

\begin{figure}[h!]
    \centering
    \includegraphics[width=0.45\textwidth]{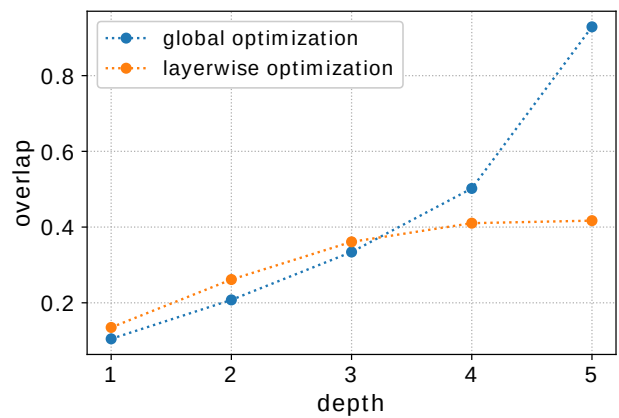}
    \caption{Overlap comparison between layerwise and global training at every layer. While global optimization aims at overall overlap improvement, the layerwise approach maximally increases overlap from layer to layer.}
    \label{lw_vs_gl}
\end{figure}

In other words, comparing overlaps at each layer/depth, layerwise optimization may achieve better overlaps than global optimization for a few initial layers, hence one observes intermediate greedy layerwise gains. However, such a strategy becomes inferior as greedy gains stagnate, see Fig. \ref{lw_vs_gl}. 

\section{Necessary saturation conditions}
\label{nec_cond}
To understand the structure of candidate states prepared by layerwise optimisation and to derive conditions necessary for saturation we recall the following:
\begin{definition}[Symmetric subspace]
 $H_s=Span\{\ket{\psi}: P_{ij}\ket{\psi}=\ket{\psi}\}$, where $P_{ij}$ is a permutation of arbitrary qubits $i$ and $j$.  
\end{definition}

\begin{definition}[Dicke basis vectors]
$\ket{e_k} = \dfrac{1}{\sqrt{C_n^k}}\sum\limits_{nontrivial} P_{ij}\ket{11...1,0,...0}$ with $k$ ones and $n-k$ zeros.
\end{definition}
Dicke state $\ket{e_k}$ is 
a uniform superposition of all $n$-qubit states with Hamming weight $k$, (a.k.a. $1$-norm $k$). These vectors form a $n+1$ dimensional orthonormal basis of $H_s$.

Symmetric subspace $H_s$ plays a key role in our consideration: since $H_x$ commutes with any qubit permutation $P_{ij}$ and $\ket{0}^{\otimes n}\equiv\ket{e_0},\ket{+}\in H_s$, then all QAOA ansatz states prepared as \eqref{ansatz} with $\ket{\bm t}=\ket{0}^{\otimes n}$ belong to the symmetric subspace.

Let us denote a state prepared with optimized $c$-depth QAOA circuit as $\ket{\psi_c} \equiv \ket{\psi_c(\bm{\gamma}^\#, \bm{\beta}^\#)}$. The ansatz states of different depth circuits are related recursively as $\ket{\psi_c} = e^{-i\beta_c H_x}e^{-i\gamma_c \ketbra{\bm t}{\bm t}}\ket{\psi_{c-1}}$ with the initial state $\ket{\psi_0}=\ket{+}^{\otimes n}$. Since $\ket{\psi_c} \in H_s$, one can expand over the Dicke basis to obtain $n+1$ components $A_k^c=\braket{e_k}{\psi_c}$.

The optimization task of layerwise training is to maximize the overlap at each depth $c$. This overlap can be written in terms of wave function coefficients $A^{c-1}_k$ of the previous layer:

\begin{align}
g_c
&\equiv\braket{0}{\psi_c}=
g_{c-1}\cos^n\beta_c e^{-i\gamma_c}+\nonumber\\
&+\sum\limits_{k=1}^n(\cos\beta_c)^{n-k}(i \sin\beta_c)^k A_k^{c-1}\sqrt{C_n^k},
\label{ll_overlap}
\end{align}
(notice that $g_{c-1}\equiv A_0^{c-1}$).

All the coefficients $A_k^{c-1}$ are functions of $2(c-1)$ variational parameters of the previous layers. In layerwise optimization these values are already fixed from previous training iterations, thus leaving $g_c$ to be a function of only current layer parameters $\beta_c$ and $\gamma_c$. 

Optimization of expression \eqref{ll_overlap} over the region of $\gamma_c$ can easily be performed using a geometrical fact that for any complex numbers $A$ and $B$, $\max\limits_\gamma |A e^{-i\gamma}+B| = |A|+|B|$:

\begin{align}
g(\beta_c) &\stackrel{\rm{def}}{=} \max\limits_{\gamma_c}|g_c|=\cos^n\beta_c(|g_{c-1}| +\nonumber\\
&+|\sum\limits_{k=1}^n(-i \tan\beta_c)^k A_k^{c-1}\sqrt{C_n^k}|),   \label{max_gamma}
\end{align}

The above expression should be supplemented with the iterative expression for amplitudes $A_k^{c-1}$. One should then calculate the action of mixer $e^{-i \beta H_x}$ on the Dicke basis vectors, which implies representation of the Hadamard transformation in the Dicke basis. However, the result of this procedure is messy and further solution requires a consideration of the general optimization problem. 

Expression \eqref{max_gamma} allows for derivation of necessary conditions for observing saturation at some depth. Assuming that saturation already happened at depth $c$ we study the structure of the output state (coefficients $A_k^{c-1}$). Saturation implies that the circuit output state cannot be trained further and optimization of additional layer parameters only returns trivial values of $\beta$. Thus, \eqref{max_gamma} should have a global maximum at position $\beta_c = 0$. We calculate derivatives $\dfrac{dg}{d\beta_c}(0)=\sqrt{C_n^1}|A_1^{l-1}|$ and $\dfrac{d^2 g}{d\beta_c^2}(0) = -n |g_{c-1}| + \sqrt{C_n^2}|A_2^{c-1}|$ and use standard conditions for maxima ($g'(0)=0$, $g''(0)\le0$) to finally arrive at
\begin{prop}[Necessary saturation conditions]
\label{prop_necesary_conditions}
States $\ket{\psi}\in H_s$ that cannot be trained by QAOA with target state $\ket{0}^{\otimes n}$ necessarily satisfy
\begin{align}
     A_1=\braket{e_1}{\psi} \propto \bra{0}^{\otimes n}H_x\ket{\psi}\propto\bra{0}^{\otimes n}X\ket{\psi}=0 ,
     \label{firt_cond}
     \\
     |A_2|=|\braket{e_2}{\psi}|\le\sqrt{\dfrac{2n}{n-1}}|\bra{0}^{\otimes n}\ket{\psi}|.
     \label{sec_cond}
\end{align}
\end{prop}

Physically, \eqref{firt_cond} and \eqref{sec_cond} imply that states $\ket{e_1}$ and $\ket{e_2}$ (if their amplitudes in $\ket{\psi}$ are large enough) are the source of state $\ket{0}^{\otimes n}$ under the action of QAOA: probability flows from states $\ket{e_{1,2}}$ to the target state $\ket{0}^{\otimes n}$.
If training saturates, this sources should be drained, as formulated in conditions \eqref{firt_cond} and \eqref{sec_cond}. 
\begin{figure}[!tbh]
   \centerline{\includegraphics[clip=true,width=3.2in]{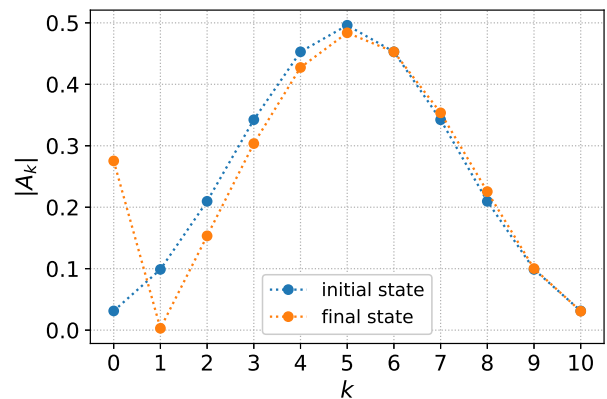}}
       \caption{Coefficients $A_k$ for $k=0,\dots,n$ with $n=10$ qubits. Blue points represent the initial state $\ket{+}^{\otimes n}$, while the orange ones the output state of a $p=10$ layerwise trained QAOA circuit. From numerical consideration we know this output state is already saturated (QAOA can't improve its overlap further.)
       QAOA performance is clearly seen as increase of the first coefficient $|A_0|$.}
 \label{coef}
 \end{figure}
Fig. \ref{coef} illustrates that when one observes saturation, the circuit output state indeed satisfies necessary  saturation conditions \eqref{firt_cond} and \eqref{sec_cond}, i.e. $A_1=0$, and $|A_2|$ is somewhat small compared to $|A_0|$.

If at least one of conditions ($\ref{firt_cond}$) and ($\ref{sec_cond}$) is violated, QAOA would manage to improve state overlap with $\ket{0}^{\otimes n}$. Thus, avoiding states that satisfy ($\ref{firt_cond}$) and ($\ref{sec_cond}$) is key to avoid saturation. 
We discuss the ways to violate these conditions and to assist training in the next section.

\section{Introducing variability to QAOA}
\label{sec_noise}
Saturation strongly limits the performance of layerwise optimization. It results from the particular structure that the greedy layerwise algorithm imposes on coefficients $A_k$. Deviating from this greedy strategy by violating conditions \eqref{firt_cond} and \eqref{sec_cond} and thus altering the structure imposed, one might avoid saturation and thereby improve training. The simplest approach would be to push amplitude $A_1$ up, for example, by perturbing amplitudes $A_1, ..., A_n$. This will violate condition \eqref{firt_cond}, thus non-zero amplitude $A_1$ might further be transferred to $\ket{0}^{\otimes n}$, removing saturation at the current layer.

\begin{figure}
   \centering
    \includegraphics[width=0.5\textwidth]{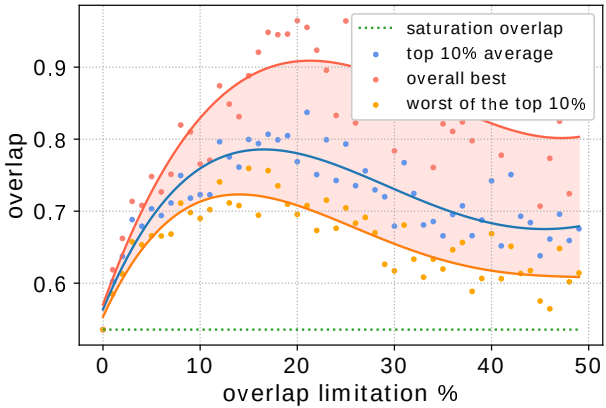}
    \caption{State preparation overlap for a 4 qubit layerwise $p=4$ QAOA versus the limitation of the maximum overlap at each layer. Polynomial fittings illustrate the best, average and worst of the top $10\%$ overlaps.}
    \label{overlap_limitation}
\end{figure}

A more algorithmic approach, i.e., to perturb the coefficient structure and thus improve layerwise training, is to limit QAOA performance on each individual layer. Limiting perfect optimization (e.g.~inducing cutoff at $90\%$ of maximal improvement at the current layer) necessarily perturbs the amplitudes and thus avoids saturation. This effect is illustrated in Fig. \ref{overlap_limitation}. Notice that threshold $p^*=n$ is now removed, allowing overlap to increase with circuit depth. 


Layerwise optimization need not preserve candidate states in the symmetric subspace $H_s$ when systematic errors (e.g., phase noise or bit-flipping) are considered. Here, \eqref{firt_cond} and \eqref{sec_cond} are no longer applicable and therefore saturation may be avoided. On one hand low probability coherent phase noise could reduce overlap by just a fraction, thus allowing further training. But on the other hand, bit flip noise would radically alter the state amplitudes thus hampering training. Indeed these are confirmed by our numerical simulations.     

We layerwise trained QAOA circuits of $p=n$ with a gradient free optimizer in the presence of coherent phase noise described as: 
\begin{equation}
   S(\phi)=\ketbra{0}{0}+e^{i\phi}\ketbra{1}{1}
\end{equation}
where $\phi$ is sampled from a normal distribution centered around 0 with variance 1. The noise is applied  with a probability $P$ to any qubit after each gate application. Experimentally, this type of noise engineering has been used to test the limits of variational algorithms in identifying and quantifying quantum phase transitions with noisy qubits  \cite{borzenkova2021variational,pechen2011engineering}. In Fig.~\ref{noise} we plot the top $10\%$ overlaps obtained in 100 trials versus the probability of applying noise. The curves are polynomial fits for the best, average and worst of the top $10\%$ overlaps.

For a trial emulation of the algorithm, we can expect a $10\%$ probability of obtaining an overlap inside the red shaded area, which for some values of $P$ are significantly higher than the overlap obtained by training layerwise in an ideal system as illustrated by the dotted green line.

Moreover, the overlap can further be increased with the addition of more layers since we are no longer limited by saturation at $p=n$. The use of a constant probability of noise may not be optimal to obtain the maximum possible overlap, but it is enough to show that its presence can offer an advantage over noiseless systems in certain scenarios.

\section{Discussion}

The formulated necessary saturation conditions \eqref{firt_cond} and \eqref{sec_cond}, in general, may not be sufficient for observing saturation since the amplitudes: $A_3, ..., A_n$ also are a source for $A_0$ under QAOA. In other words, the necessary saturation conditions ensure that overlap function \eqref{max_gamma} has a maximum at $\beta_c=0$, which still may just be a local maximum. The absolute maximum for layerwise training, searched and found by the optimizer, can be located in the middle of parameter region, thus could not be captured by conditions \eqref{firt_cond} and \eqref{sec_cond}.

\begin{figure}[h]
    \centering
    \includegraphics[width=0.45\textwidth]{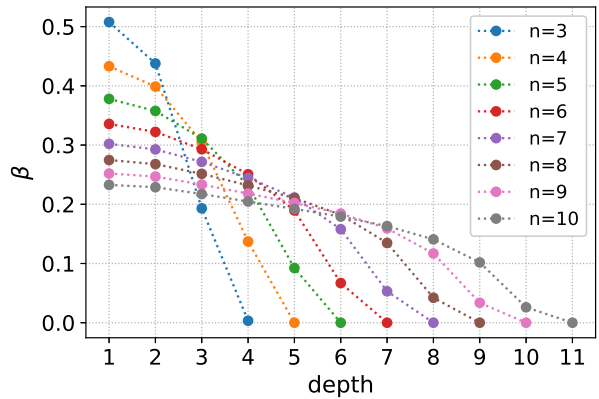}
    \caption{Optimal $\bm{\beta}$ for depths $p=n+1$. For each case, the last layer is affected by saturability, thus returning a trivial optimal angle for $\beta_p$.  }
    \label{betas}
\end{figure}

However, based on the values of optimal parameters obtained numerically (Fig. \ref{betas}) for large $n$ we see not only that optimal $\beta$s are small, they decrease with increasing depth. As a result, global maximum of the overlap shifts towards smaller values of $\beta$, and it finally stagnates at $\beta=0$ (causing saturation).  
Thus we conclude that the formulated conditions for maximum overlap at $\beta_c=0$ \eqref{firt_cond} and \eqref{sec_cond} are not only necessary for observing saturation, but also sufficient provided that the optimal angles are small. 

For large $n$ it can be understood as follows:
It was shown in \cite{akshay2021parameter} that first layer optimal parameter $\beta_1\approx\dfrac{\pi}{n}$ is small. That implies, the first layer marginally affects amplitudes $A_k$ for large $k$, leaving the $\ket{+}^{\otimes n}$ state almost unmodified up to a global phase. Therefore, optimal parameters at subsequent layers also remain small, which again preserve $\ket{+}^{\otimes n}$ state. Thus in the output state only the lowest amplitudes are modified, while the others stay almost the same, as illustrated in Fig. \ref{coef}.
In other words, layerwise QAOA manages to modify only slightly the initial state, keeping the large amplitudes almost preserved, causing it to saturate.

While we formulated necessary saturation conditions, expression \eqref{max_gamma} allows one to show (see Appendix) the existence of several families of states in the symmetric subspace for which layerwise QAOA does not allow for overlap improvement. The observed saturation effect implies that layerwise QAOA necessarily reaches one of these non-trainable states which causes saturation. 

Multiple strategies can be employed in order to avoid non-trainable states in QAOA. The perturbation to the state has to be able to violate \eqref{firt_cond} and/or \eqref{sec_cond} but not big enough to loose the amplitude $A_0$ that accumulates with every layer, illustrated in Fig. \ref{coef}. Noise of the form $e^{-i\sigma \phi}$, where $\sigma$ is a Pauli matrix and  $\langle\phi\rangle=0$ , makes small changes to the inner product for a state $\ket{\psi}$
\begin{equation}
    \bra{\psi}e^{-i\sigma \phi}\ket{\psi}\approx\bra{\psi}\left(1-i\sigma\phi-\frac{\phi^2}{2}\right)\ket{\psi}=1-\frac{\langle\phi^2\rangle}{2}
\end{equation}

In contrast, bit flip noise does not have a good representation as a series expansion, thus altering the structure of the state in a way that harms the training.

\appendix

\section{Non-trainable states}
From expression (\ref{max_gamma}) also allows to see, that in the symmetric subspace there exist some states different from $\ket{0}^{\otimes n}$ that cannot be trained by QAOA to improve the overlap with $\ket{0}^{\otimes n}$. For instance, it is easy to see that if one tries to implement QAOA on the state 

\begin{equation}
    \ket{\psi}=A_0\ket{0}+A_2\ket{e_2},
    \label{nt_state}
\end{equation}
with $|A_2|\le \dfrac{|A_0|}{\sqrt{C_n^2}}$, the resulting overlap (\ref{max_gamma}) would be

\begin{equation}
    g=(|A_0|-\sqrt{C_n^2} |A_2|)\cos^n\beta+\sqrt{C_n^2}|A_2|\cos^{n-2}\beta,
\end{equation}
with maximum at $\beta=0$. Therefore, QAOA is not able to improve overlap of state (\ref{nt_state}) with $\ket{0}$.

\begin{widetext}
\onecolumngrid
\begin{figure}[h]
\begin{minipage}[h]{0.47\linewidth}
\center{\includegraphics[width=1\linewidth]{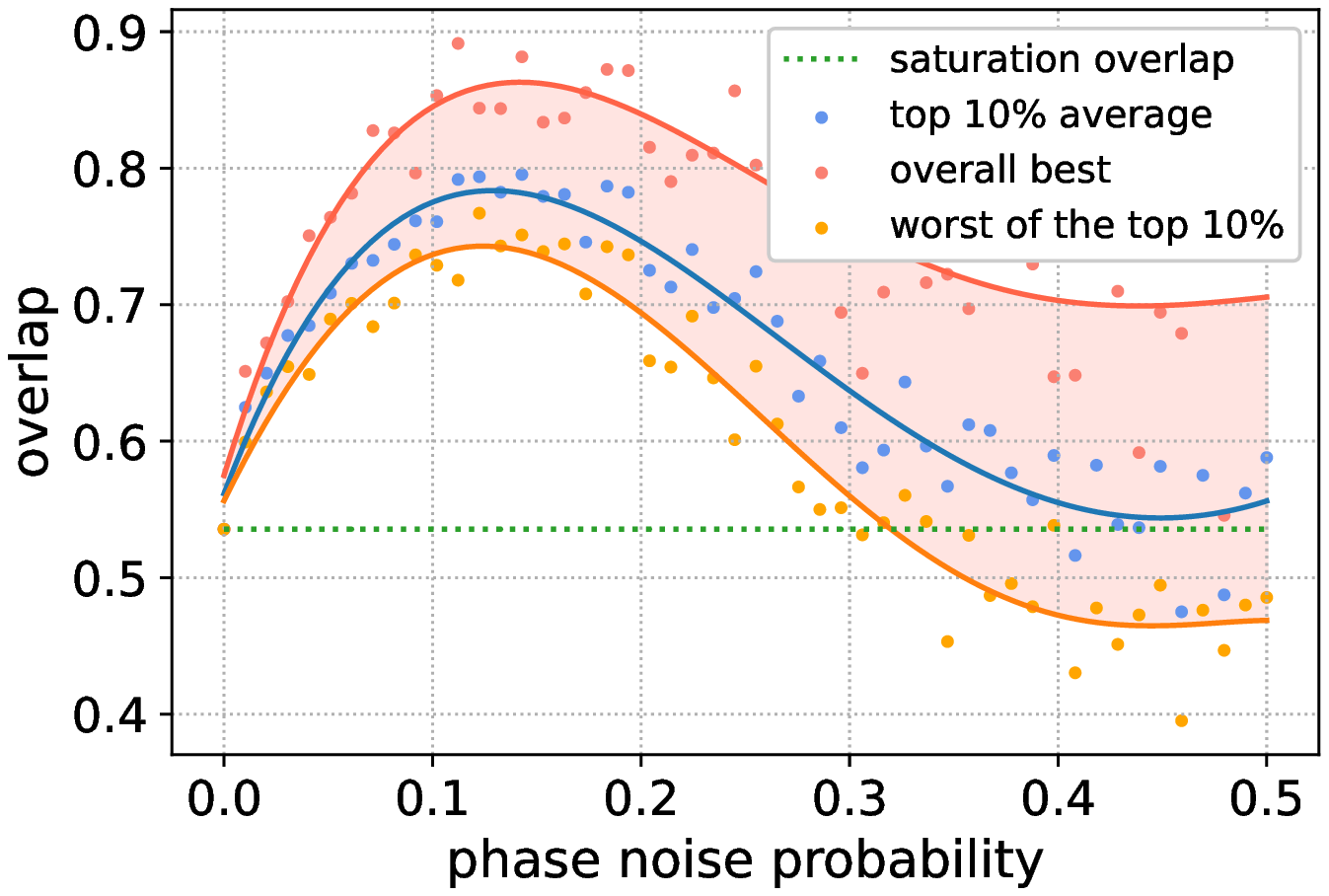}}     \phantom{.}~~~~~~$(n=4)$ \\
\end{minipage}
\hfill
\begin{minipage}[h]{0.47\linewidth}
\center{\includegraphics[width=1\linewidth]{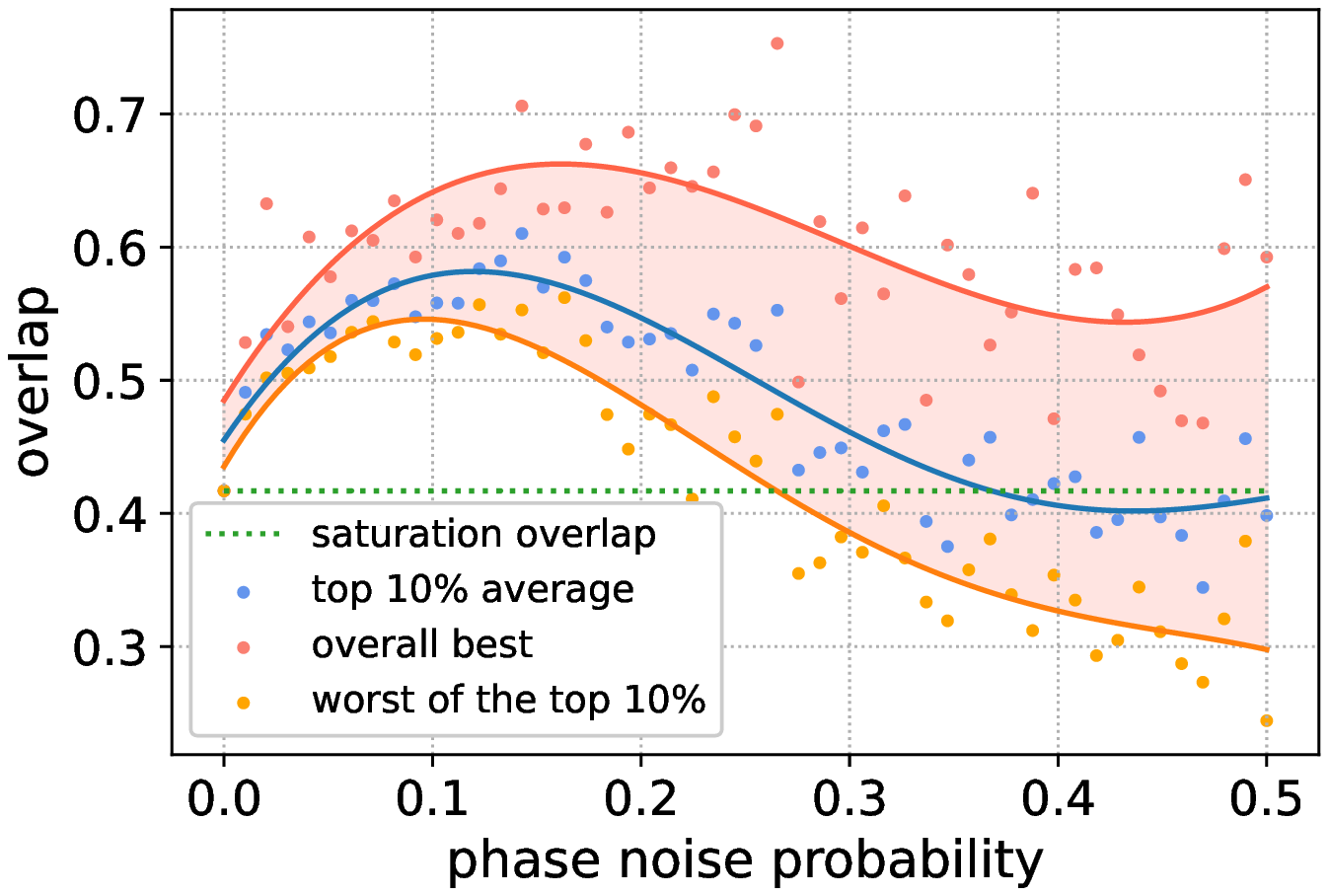}}  \phantom{.}~~~~~~$(n=5)$\\
\end{minipage}
\vfill
\begin{minipage}[h]{0.47\linewidth}
\center{\includegraphics[width=1\linewidth]{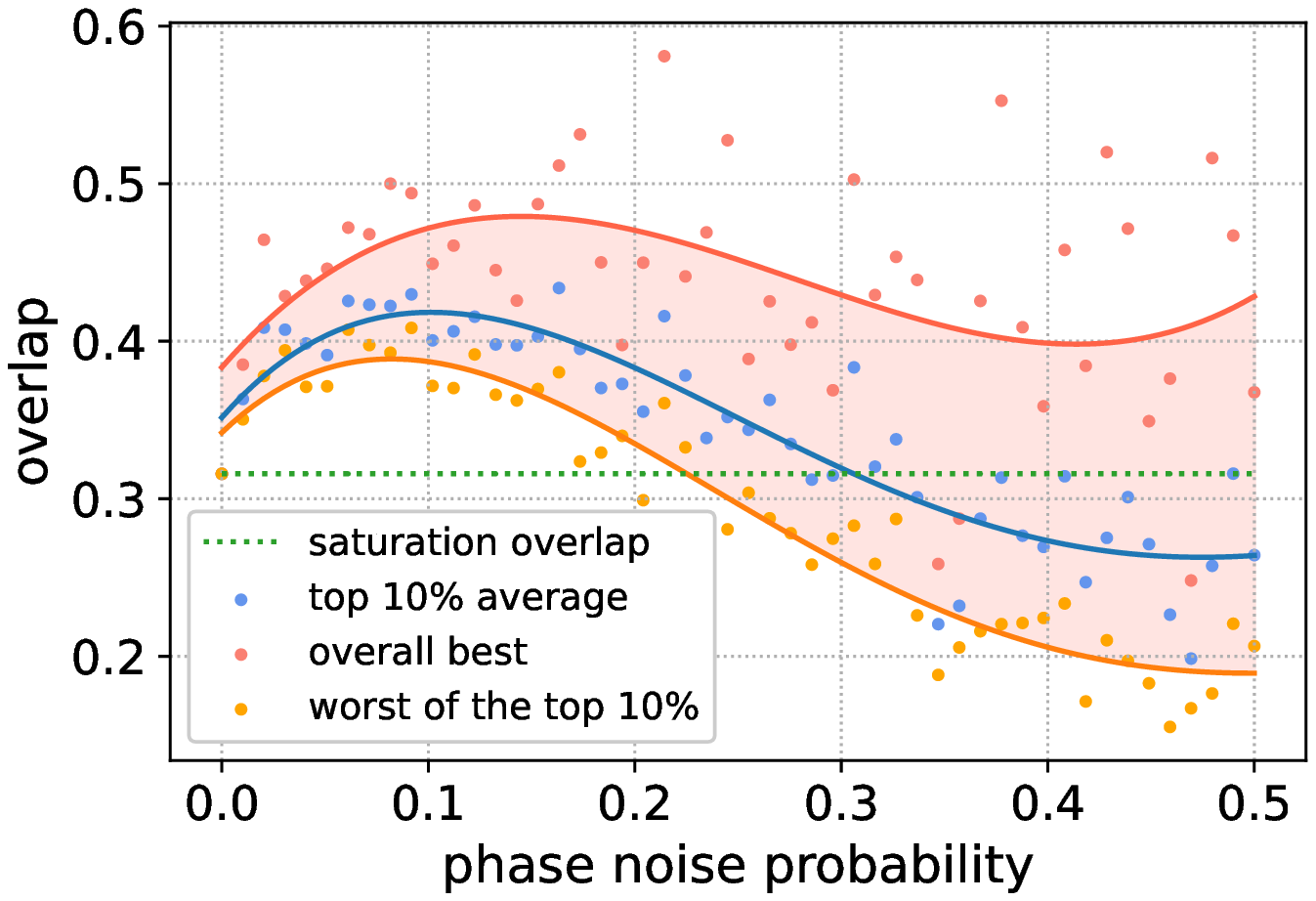}}  \phantom{.}~~~~~~~$(n=6)$ \\
\end{minipage}
\hfill
\begin{minipage}[h]{0.47\linewidth}
\center{\includegraphics[width=1\linewidth]{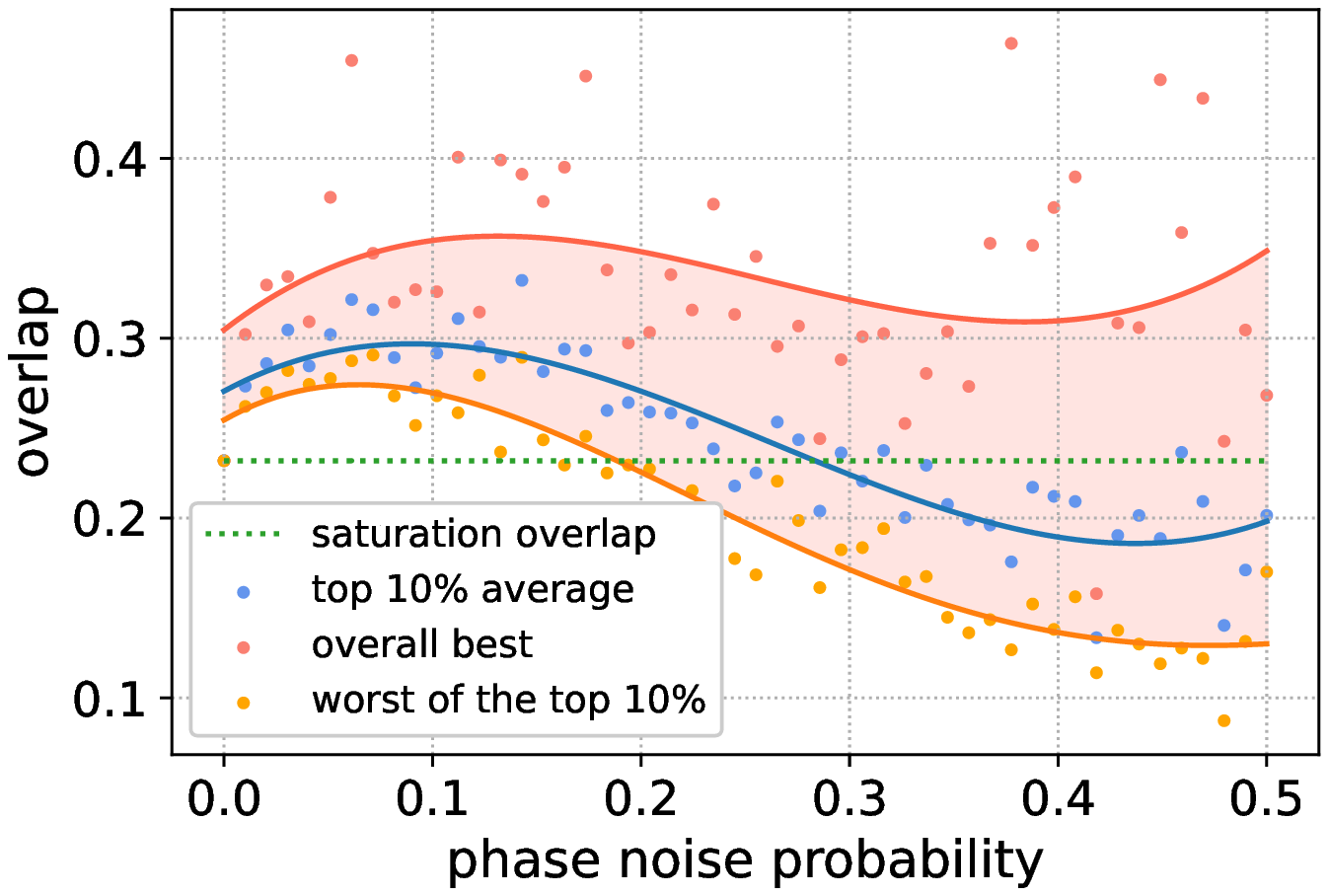}}  \phantom{.}~~~~~~$(n=7)$ \\
\end{minipage}\caption{State preparation overlap of layerwise QAOA in the presence of phase noise for $p=n$. The phase noise angle is sampled from a normal distribution centered at 0 with variance 1. For a certain probability of noise, a run of the algorithm will have a $10\%$ probability of overlap inside the shaded red region, considerably higher than the noiseless overlap illustrated as a dotted green line.}
\label{noise}
\end{figure}
\end{widetext}

\bibliography{refs.bib}

\begin{thebibliography}{10}

\bibitem{harrigan2021quantum}
Matthew~P Harrigan, Kevin~J Sung, Matthew Neeley, Kevin~J Satzinger, Frank
  Arute, Kunal Arya, Juan Atalaya, Joseph~C Bardin, Rami Barends, Sergio Boixo,
  et~al.
\newblock Quantum approximate optimization of non-planar graph problems on a
  planar superconducting processor.
\newblock {\em Nature Physics}, 17(3):332--336, 2021.

\bibitem{pagano2019quantum}
Guido Pagano, Aniruddha Bapat, Patrick Becker, Katherine~S. Collins, Arinjoy
  De, Paul~W. Hess, Harvey~B. Kaplan, Antonis Kyprianidis, Wen~Lin Tan,
  Christopher Baldwin, Lucas~T. Brady, Abhinav Deshpande, Fangli Liu, Stephen
  Jordan, Alexey~V. Gorshkov, and Christopher Monroe.
\newblock Quantum approximate optimization of the long-range ising model with a
  trapped-ion quantum simulator.
\newblock {\em Proceedings of the National Academy of Sciences},
  117(41):25396--25401, 2020.

\bibitem{guerreschi2019qaoa}
Gian~Giacomo Guerreschi and Anne~Y Matsuura.
\newblock Qaoa for max-cut requires hundreds of qubits for quantum speed-up.
\newblock {\em Scientific reports}, 9(1):1--7, 2019.

\bibitem{butko2020understanding}
Anastasiia Butko, George Michelogiannakis, Samuel Williams, Costin Iancu, David
  Donofrio, John Shalf, Jonathan Carter, and Irfan Siddiqi.
\newblock Understanding quantum control processor capabilities and limitations
  through circuit characterization.
\newblock In {\em 2020 International Conference on Rebooting Computing (ICRC)},
  pages 66--75, 2020.

\bibitem{Farhi2014}
Edward Farhi, Jeffrey Goldstone, and Sam Gutmann.
\newblock A quantum approximate optimization algorithm.
\newblock {\em arXiv preprint arXiv:1411.4028}, 2014.

\bibitem{niu2019optimizing}
Murphy~Yuezhen Niu, Sirui Lu, and Isaac~L Chuang.
\newblock Optimizing qaoa: Success probability and runtime dependence on
  circuit depth.
\newblock {\em arXiv preprint arXiv:1905.12134}, May 2019.

\bibitem{lloyd2018quantum}
Seth Lloyd.
\newblock Quantum approximate optimization is computationally universal.
\newblock {\em arXiv preprint arXiv:1812.11075}, 2018.

\bibitem{morales2020universality}
Mauro~ES Morales, JD~Biamonte, and Zolt{\'a}n Zimbor{\'a}s.
\newblock On the universality of the quantum approximate optimization
  algorithm.
\newblock {\em Quantum Information Processing}, 19(9):1--26, 2020.

\bibitem{Zhou2020}
Leo Zhou, Sheng-Tao Wang, Soonwon Choi, Hannes Pichler, and Mikhail~D. Lukin.
\newblock Quantum approximate optimization algorithm: Performance, mechanism,
  and implementation on near-term devices.
\newblock {\em Phys. Rev. X}, 10:021067, Jun 2020.

\bibitem{wang2020x}
Zhihui Wang, Nicholas~C Rubin, Jason~M Dominy, and Eleanor~G Rieffel.
\newblock X y mixers: Analytical and numerical results for the quantum
  alternating operator ansatz.
\newblock {\em Physical Review A}, 101(1):012320, 2020.

\bibitem{Brady2021}
Lucas~T. Brady, Christopher~L. Baldwin, Aniruddha Bapat, Yaroslav Kharkov, and
  Alexey~V. Gorshkov.
\newblock {Optimal Protocols in Quantum Annealing and Quantum Approximate
  Optimization Algorithm Problems}.
\newblock {\em Physical Review Letters}, 126(7):070505, Feb 2021.

\bibitem{Farhi2016}
Edward Farhi and Aram~W Harrow.
\newblock Quantum supremacy through the quantum approximate optimization
  algorithm.
\newblock {\em arXiv preprint arXiv:1602.07674}, 2016.

\bibitem{Akshay2020}
V.~Akshay, H.~Philathong, M.~E.S. Morales, and J.~D. Biamonte.
\newblock {Reachability Deficits in Quantum Approximate Optimization}.
\newblock {\em Physical Review Letters}, 124(9):090504, Mar 2020.

\bibitem{Farhi2019a}
Edward Farhi, Jeffrey Goldstone, Sam Gutmann, and Leo Zhou.
\newblock The quantum approximate optimization algorithm and the
  sherrington-kirkpatrick model at infinite size.
\newblock {\em arXiv preprint arXiv:1910.08187}, Oct 2019.

\bibitem{Wauters2020}
Matteo~M Wauters, Glen Bigan~Mbeng, and Giuseppe~E Santoro.
\newblock Polynomial scaling of qaoa for ground-state preparation of the
  fully-connected p-spin ferromagnet.
\newblock {\em arXiv e-prints}, pages arXiv--2003, 2020.

\bibitem{Claes2021}
Jahan Claes and Wim van Dam.
\newblock Instance independence of single layer quantum approximate
  optimization algorithm on mixed-spin models at infinite size.
\newblock {\em arXiv preprint arXiv:2102.12043}, 2021.

\bibitem{Zhou}
Leo Zhou, Sheng-Tao Wang, Soonwon Choi, Hannes Pichler, and Mikhail~D. Lukin.
\newblock Quantum approximate optimization algorithm: Performance, mechanism,
  and implementation on near-term devices.
\newblock {\em Phys. Rev. X}, 10:021067, Jun 2020.

\bibitem{akshay2021parameter}
V~Akshay, D~Rabinovich, E~Campos, and J~Biamonte.
\newblock Parameter concentration in quantum approximate optimization.
\newblock {\em arXiv preprint arXiv:2103.11976}, 2021.

\bibitem{Jiang2017a}
Zhang Jiang, Eleanor~G. Rieffel, and Zhihui Wang.
\newblock {Near-optimal quantum circuit for Grover's unstructured search using
  a transverse field}.
\newblock {\em Physical Review A}, 95(6), Feb 2017.

\bibitem{hastings2019classical}
Matthew~B Hastings.
\newblock Classical and quantum bounded depth approximation algorithms.
\newblock {\em arXiv preprint arXiv:1905.07047}, May 2019.

\bibitem{Bravyi2019}
Sergey Bravyi, Alexander Kliesch, Robert Koenig, and Eugene Tang.
\newblock {Obstacles to State Preparation and Variational Optimization from
  Symmetry Protection}.
\newblock {\em Physical Review Letters}, 125(26), Oct 2019.

\bibitem{shaydulin2021exploiting}
Ruslan Shaydulin and Stefan~M Wild.
\newblock Exploiting symmetry reduces the cost of training qaoa.
\newblock {\em IEEE Transactions on Quantum Engineering}, 2:1--9, 2021.

\bibitem{streif2019comparison}
Michael Streif and Martin Leib.
\newblock Comparison of qaoa with quantum and simulated annealing.
\newblock {\em arXiv preprint arXiv:1901.01903}, Jan 2019.

\bibitem{skolik2021layerwise}
Andrea Skolik, Jarrod~R McClean, Masoud Mohseni, Patrick van~der Smagt, and
  Martin Leib.
\newblock Layerwise learning for quantum neural networks.
\newblock {\em Quantum Machine Intelligence}, 3(1):1--11, 2021.

\bibitem{campos2020abrupt}
Ernesto Campos, Aly Nasrallah, and Jacob Biamonte.
\newblock Abrupt transitions in variational quantum circuit training.
\newblock {\em Phys. Rev. A}, 103:032607, Mar 2021.

\bibitem{bravyi2018correcting}
Sergey Bravyi, Matthias Englbrecht, Robert K{\"o}nig, and Nolan Peard.
\newblock Correcting coherent errors with surface codes.
\newblock {\em npj Quantum Information}, 4(1):1--6, 2018.

\bibitem{sharma2020noise}
Kunal Sharma, Sumeet Khatri, Marco Cerezo, and Patrick~J Coles.
\newblock Noise resilience of variational quantum compiling.
\newblock {\em New Journal of Physics}, 22(4):043006, 2020.

\bibitem{gentini2020noise}
Laura Gentini, Alessandro Cuccoli, Stefano Pirandola, Paola Verrucchi, and
  Leonardo Banchi.
\newblock Noise-resilient variational hybrid quantum-classical optimization.
\newblock {\em Physical Review A}, 102(5):052414, 2020.

\bibitem{cincio2021machine}
Lukasz Cincio, Kenneth Rudinger, Mohan Sarovar, and Patrick~J Coles.
\newblock Machine learning of noise-resilient quantum circuits.
\newblock {\em PRX Quantum}, 2(1):010324, 2021.

\bibitem{mcclean2016theory}
Jarrod~R McClean, Jonathan Romero, Ryan Babbush, and Al{\'a}n Aspuru-Guzik.
\newblock The theory of variational hybrid quantum-classical algorithms.
\newblock {\em New Journal of Physics}, 18(2):023023, 2016.

\bibitem{mcclean2017hybrid}
Jarrod~R McClean, Mollie~E Kimchi-Schwartz, Jonathan Carter, and Wibe~A
  De~Jong.
\newblock Hybrid quantum-classical hierarchy for mitigation of decoherence and
  determination of excited states.
\newblock {\em Physical Review A}, 95(4):042308, 2017.

\bibitem{cao2021noise}
Chenfeng Cao and Xin Wang.
\newblock Noise-assisted quantum autoencoder.
\newblock {\em Physical Review Applied}, 15(5):054012, 2021.

\bibitem{borzenkova2021variational}
OV~Borzenkova, GI~Struchalin, AS~Kardashin, VV~Krasnikov, NN~Skryabin,
  SS~Straupe, SP~Kulik, and JD~Biamonte.
\newblock Variational simulation of schwinger's hamiltonian with polarization
  qubits.
\newblock {\em Applied Physics Letters}, 118(14):144002, 2021.

\bibitem{pechen2011engineering}
Alexander Pechen.
\newblock Engineering arbitrary pure and mixed quantum states.
\newblock {\em Physical Review A}, 84(4):042106, 2011.

\end{thebibliography}
\bibliographystyle{unsrt}
\end{document}